\documentclass{elsart3p}

\usepackage{natbib}
\usepackage{graphicx}
\usepackage{amssymb}

\def\msol{M_\odot}
\def\mdens{\rm g~cm^{-3}}

\begin{document}
\begin{frontmatter}



\title{Fast rotation of neutron stars
       and equation of state of dense matter}

\author{Pawe{\l} Haensel\corauthref{cor1}}
\author{Julian L. Zdunik and}
\author{Micha{\l} Bejger}

\address{Nicolaus Copernicus Astronomical Center, Polish
           Academy of Sciences, Bartycka 18, PL-00-716 Warszawa, Poland}

\corauth[cor1]{haensel@camk.edu.pl}

\begin{abstract}
Fast rotation of compact stars (at submillisecond period)
and, in particular, their stability,
are sensitive to the equation of state (EOS) of dense matter.
Recent observations of XTE J1739-285 suggest that
it contains a neutron star rotating at 1122 Hz \citep{Kaaret2007}.
At such rotational frequency  the effects of rotation
on star's structure are significant.
We study the interplay of fast rotation, EOS, and
gravitational mass of a submillisecond pulsar. We
discuss the EOS dependence of spin-up to a submillisecond period,
via mass accretion from a disk in a low-mass X-ray binary.
\end{abstract}

\begin{keyword}
Dense matter \sep Equation of state \sep stars: neutron \sep
stars: rotation \sep Pulsars \sep Low-mass X-ray binaries

\PACS 26.60+c \sep 97.60.Gb \sep 97.60.Jd \sep 97.80.Jd

\end{keyword}

\end{frontmatter}

\section{Introduction}
\label{sect:introd}
Neutron stars, and more generally, compact stars (neutron
stars, hybrid hadron-quark stars, quark stars) are the
densest stellar objects in the Universe. Due to their
compactness and strong gravity, compact stars can be very fast
rotators. In the present paper we limit ourselves to a rigid
rotation. Theoretical calculations
 show that compact stars could  rotate at sub-millisecond periods
 (i.e., at frequency $f=1/{\rm
period}>$1000 Hz: \citealt{CST1994,Salgado1994}).

The quest for fast rotating compact stars has an interesting
history. The first millisecond pulsar B1937+21, rotating at
$f=641$ Hz \citep{Backer1982}, remained the most rapid one for
24 years. In January 2006, discovery of a faster pulsar
J1748-2446ad rotating at $f=716$ Hz was announced
\citep{Hessels2006}. However, sub-kHz frequencies are still
too low to significantly affect the structure of massive
neutron stars with $M>1M_\odot$ \citep{STW1983,NSB1}. Actually,
pulsars B1937+21 and J1748-2446ad still rotate in  a {\it slow
rotation} regime, because their $f$ is significantly smaller
than the mass shedding (Keplerian) frequency $f_{\rm K}$. In
the slow rotation regime rotational effects in neutron star
structure, e.g., polar flattening, are $\propto (f/f_{\rm
K})^2\ll 1$, and hence not large. Rapid rotation regime for
$M>1M_\odot$ requires sub-millisecond pulsars with supra-kHz
frequencies ($f>1000$ Hz).

Exciting news came in  December 2006.
\citet{Kaaret2007} reported a discovery of oscillation
frequency $f=1122$ Hz in an X-ray burst from the X-ray
transient,  XTE J1739-285. \citet{Kaaret2007}
wrote cautiously "this oscillation frequency suggests that XTE
J1739-285 contains the fastest rotating neutron star yet
found". If confirmed, this would be the first detection of a
sub-millisecond pulsar (discovery of a 0.5 ms pulsar in
SN1987A remnant, announced in January 1989, was withdrawn one
year later).

Fast rotation of compact stars is sensitive to the stellar
mass and to the equation of state (EOS). Hydrostatic,
stationary configurations of neutron stars rotating at given
rotation frequency $f$ form a one-parameter family, labeled by
the central density. This family - a curve in the mass -
equatorial radius plane - is limited by two instabilities. On
the high central density side, it is instability with respect
to axi-symmetric perturbations, making the star collapse into
a Kerr black hole. The low central density boundary results
from the mass shedding from the equator.  In the present paper
we discuss the dependence of rotation at $f>1000$ Hz on the
poorly known  EOS, and derive constraints on the EOS of
neutron stars which could result from future observations of
stably rotating sub-millisecond pulsars.

The plan of the paper is as follows. In Sect.\ 1 we briefly
describe EOSs used in our calculations. Numerical methods of
solving equations of hydrostatic equilibrium of rigidly
rotating compact stars in General Relativity, and criteria for
their stability, are presented in Sect.\ 2. In Sect.\ 3 we
consider the EOS dependence case of rotation at $f=1122$ Hz. A
systematic study of the  EOS-dependence of  fast rotation at
$f=1000-1600$ Hz is presented in Sect.\ 4. Reaching fast
rotation via disk accretion  in LMXBs and the EOS-dependence
of the spin-up track is reviewed in Sect.\ 5. Finally, Sect.\
6 summarizes briefly main results reported in  the paper.

We are pleased to present our paper on fast rotation of
neutron stars in the proceedings of the conference in honor of
J.-P. Lasota. In 1990s Jean-Pierre became intrigued by a
puzzling precision of ``empirical formula'' for the absolute
upper bound on rotation frequency of compact stars, proposed
by two of us (PH and JLZ) on the aftermath of ill fated
``discovery'' of 0.5 ms pulsar in SN 1987A \citep{HZ1989}. Two
joint papers resulted from our collaboration
\citep{LHA1996,HLZ1999}, some progress has been made, but the
basic puzzle still remains to be solved.

\section{Equations of state}
\label{sect:EOSs}
\begin{figure}[h]
\includegraphics[width=\columnwidth]{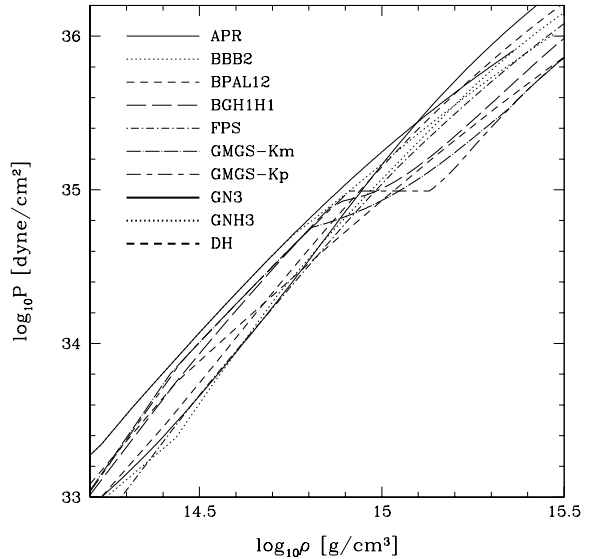}
\caption{The equations of state in the ${\rm log}P - {\rm
log}\rho$ plane. For labels - see Table 1.}
\label{fig:eos}
\end{figure}

 In view of a high degree of our ignorance of EOS of dense
 matter at supranuclear densities ($\rho>3\times 10^{14}~
 \mdens$) we considered a broad set of theoretical models.
The set of ten equations of state (EOSs) considered in the
paper is presented in Fig.~\ref{fig:eos}. The EOSs are also
listed in Table 1, where the basic informations (label of an
EOS, theory of dense matter, reference to the original paper)
are also collected.

Two EOSs were chosen to represent a soft (BPAL12) and stiff
(GN3) extreme cases. These two extreme EOSs should not be
considered as ``realistic'', but they are used just to
``bound'' the models from the soft and the stiff sides.

Four EOSs are based on realistic models involving only
nucleons (FPS, BBB2, DH, APR). The next four  EOSs  are
softened at high  density either by the appearance of hyperons
(GNH3, BGN1H1), or a phase transition (GMGS-Km, GMGS-Kp). A
softening in the latter case is clearly visible in
Fig.~\ref{fig:eos} at pressure $P\sim 10^{35}~{\rm dyn/cm^2}$.

Two EOSs, GMGS-Kp and GMGS-Km,  describe nucleon matter with a
first order phase transition  to a kaon condensed state. In
both cases the hadronic Lagrangian is the same. However, to
get GMGS-Kp we assumed that the phase transition takes place
between two pure phases and is accompanied by a density jump,
calculated using the Maxwell construction. The  GMGS-Km EOS
was obtained  assuming that the transition occurs via a mixed
state of two phases (Gibbs construction). A mixed state is
energetically preferred  when the surface tension between the
two phases is below a certain critical value. As the value of
the surface tension is very uncertain, we considered both
cases.

\begin{table*}
{{\bf Table 1} Equations of state. N - nucleons and leptons
only.  NH - nucleons, hyperons, and leptons.  Labels of EOSs
are composed from first letters of names of authors of EOSs of
the core. In all cases but FPS, the EOS of the crust is the DH
one. For FPS EOS its own crust model is used.}
\begin{center}
\begin{tabular}[t]{c|c|c}
\hline\hline EOS & model & reference\\ \hline
 BPAL12  & N energy density functional  & \citet{Bombaci1995}\\
 FPS     & N energy density functional  & \citet{ FPS1989}\\
 GN3 & N relativistic mean field   &  \citet{Glend1985}\\
 DH & N energy density functional    &  \citet{DH2001}\\
 APR   & N variational theory $^a$    &  \citet{APR1998}\\
 BGN1H1 & NH, energy density functional    &  \citet{BG1997}\\
 GNHQm2 & NH + mixed baryon-quark state  &   \citet{GlendBOOK}\\
 BBB2 & N Brueckner theory  &  \citet{BBB1997}\\
 GNH3 & NH relativistic mean field &  \citet{Glend1985}\\
 GMGS-Km & N + mixed nucleon-kaon condensed $^b$ & \citet{PonsKcond2000}\\
 GMGS-Kp & N + pure kaon condensed $^b$ & \citet{PonsKcond2000}\\
\hline\hline
\end{tabular}
\end{center}
\vskip 10pt{$^a$  A18$\delta$+UIX$^*$ model of
\citet{APR1998}.\par $^b$ GM+GS model with $U_K^{\rm
lin}=-130~$MeV.}
\end{table*}

\section{Calculation of stationary rotating configurations and their stability}
\label{sect:calculations-stability}
We computed   stationary configurations of rigidly rotating
neutron stars in the framework of General Relativity by
solving the Einstein equations for stationary axi-symmetric
spacetime \citep{BGSM1993,GourgSS1999}. Numerical computations
have been performed  using  the {\tt rotstar} code from the
LORENE library ({\tt http://www.lorene.obspm.fr}). We
calculated one-parameter families of stationary 2-D
configurations  for ten EOSs, presented
 in Fig.~\ref{fig:eos} and Table 1.

Apart from fulfilling the equations of hydrostatic
equilibrium, stationary configurations were required to be
stable. Two instabilities were considered.
\vskip 2mm
\parindent 0pt
{\it Mass-shedding.} Stability with respect to the
mass shedding from the equator implies that at a given
gravitational mass $M$ the circumferential equatorial radius
$R_{\rm eq}$ should be smaller than $R_{\rm max}$ which
corresponds to the mass shedding (Keplerian) limit.  The value
of $R_{\rm max}$ results from the condition that the frequency
of a test particle at circular equatorial orbit of radius
$R_{\rm max}$ just above the equator of the {\it actual
rotating star}  is equal to the rotational frequency of the
star. This stability condition sets the bound on our rotating
configurations from the right side on $M - R_{\rm eq}$ plane.
It fixes the largest radius and the smallest central density,
allowed for stable stationary configurations.

\vskip 2mm
 {\it Axi-symmetric oscillations.}
 Instability with respect to these oscillations determines
  the bound  for most compact stars, with  the smallest  radius
and the highest central density (i.e., from the left side on
the $M - R_{\rm eq}$ plane). This bound  is determined by
the condition:
 \begin{equation}
 \left({\partial M\over \partial \rho_{\rm c}}
 \right)_J=0~.
 \label{eq:ax-sym.stab.line}
 \end{equation}
For stable configurations:
 \begin{equation}
 \left({\partial M\over \partial \rho_{\rm c}}
 \right)_J>0~.
 \label{eq:ax-sym.stab}
 \end{equation}
In the opposite case,
 $\left({\partial M/\partial \rho_{\rm c}}
 \right)_J<0$, a compact star star is doomed to collapse into
 a Kerr black hole.
 \parindent 21pt

\section{An example: compact  stars at 1122 Hz}
\label{sect:NS-1122Hz}
In this section we present the parameters of the stellar configurations
rotating at frequency 1122~Hz, a suggested rotation frequency
of XTE J1739-285. For details and discussion see \citet{BHZ2007}.
\begin{figure}[h]
\centering
\includegraphics[width=\columnwidth]{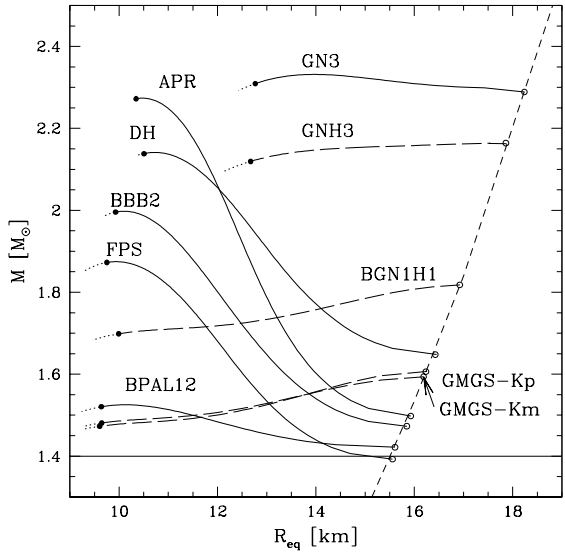}
\caption{Gravitational mass, $M$, vs. circumferential equatorial radius,
$R_\mathrm{eq}$,    for neutron stars stably rotating at $f=1122$
Hz, for ten EOSs  (Fig.~\ref{fig:eos}).
Small-radius termination by filled circle: setting-in of
instability with respect to the axi-symmetric perturbations.
Dotted segments to the left of the filled circles: configurations unstable with
respect to those perturbations.
 Large-radius termination by an open circle: the mass-shedding
instability. The mass-shedding points are very well fitted by the
dashed curve  $R_{\rm min}=15.52\;(M/1.4 M_\odot)^{1/3}\;$km.
For further explanation see the text.
 }
\label{fig:MR-NS}
\end{figure}

\begin{figure}[h]
\centering
\includegraphics[width=\columnwidth]{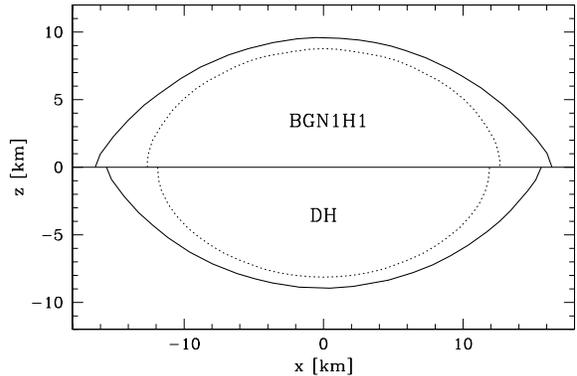}
\caption{Cross section in the plane passing through the
rotational axis of neutron stars rotating at the mass-shedding
limit  $f_{\rm K}=1122$ Hz (i.e., with $R_{\rm eq}=R_{\rm
max}$), for the BGN1H1 EOS ( $z>0$) and DH EOS ($z<0$). The
coordinates $x$ and $z$ are defined as $x=r\mathrm{sin}\theta
\mathrm{cos}\phi$, $z=r\mathrm{cos}\theta$, where $r$ is
radial coordinate in the space-time metric. Dotted contour -
crust-core boundary.
 }
\label{fig:shapesNS1}
\end{figure}

  In Fig.\  \ref{fig:MR-NS} we show the $M(R_{\rm eq})$ plots
  of stable stationary configurations rotating
  at $f=1122$ Hz. We considered EOSs  from Table 1.
  The relation between the calculated values
  of $M$ and $R_{\rm eq}$ at the "mass shedding point" is extremely
well approximated by the formula for the orbital frequency for
a test particle  orbiting at  $r=R_{\rm eq}$ in the
Schwarzschild space-time of a {\it spherical mass} $M$ (which
can be replaced by a point mass $M$ at $r=0$). Let us  denote
the orbital frequency of such a test particle by $f^{\rm
Schw.}_{\rm orb}(M,R_{\rm eq})$. The locus of points
satisfying $f^{\rm Schw.}_{\rm orb}(M,R_{\rm eq})=1122\;$Hz is
represented by a dash line in Fig.\ \ref{fig:MR-NS}. The
points on the dash line satisfy  the relation
\begin{equation}
{1\over 2\pi}\left( {GM\over{ R_{\rm eq}}^3}\right)^{1/2}=1122~{\rm
Hz}~.
\label{eq:f-orb.1122Hz}
\end{equation}
This formula,  obtained for the Schwarzschild metric,
coincides with that for  Newtonian gravity for a point mass
$M$. As one can see, it  passes through (or extremely close
to) the open circles denoting the (numerically calculated)
actual mass shedding (Keplerian) configurations. This  is a
quite remarkable property, in view of rapid rotation and
strong flattening of neutron star at the mass-shedding point,
visualized in Fig.\ \ref{fig:shapesNS1}.

\begin{figure}[h]
\centering
\includegraphics[width=\columnwidth]{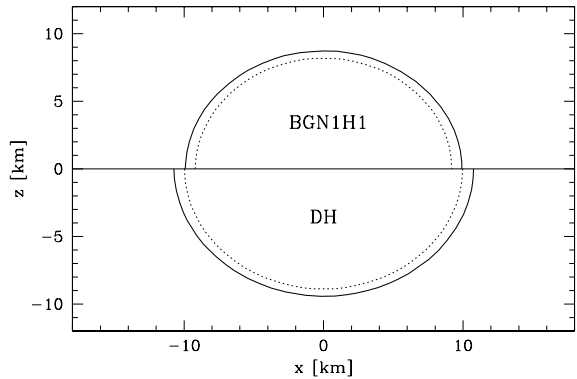}
\caption{Cross section in the plane passing through the
rotational axis of neutron stars rotating at 1122 Hz at the
axi-symmetric instability limit  (i.e., with $R_{\rm
eq}=R_{\rm min}$), for the BGN1H1 EOS ( $z>0$) and DH EOS
($z<0$). Notations as in Fig.\ \ref{fig:shapesNS1}.
 }
\label{fig:shapesNS2}
\end{figure}

Equation (\ref{eq:f-orb.1122Hz}) implies a useful constraint
for compact stars rotating at $1122$ Hz:
\begin{equation}
R_{\rm max}=15.52\;\left({M\over 1.4\;M_\odot}\right)^{1/3}\;{\rm km}~.
\label{eq:Rmax.1122Hz}
\end{equation}
%
\section{Submillisecond pulsars}
\label{sect:sub-ms}
In this section we present results for compact stars
rotating at sub-millisecond periods (supra-kHz frequencies)
for a broad range of frequencies, 1000 - 1600 Hz. As in Sect.\
\ref{sect:NS-1122Hz}, calculations are performed for ten EOSs
from Table 1.
\begin{figure}
\includegraphics[width=\columnwidth]{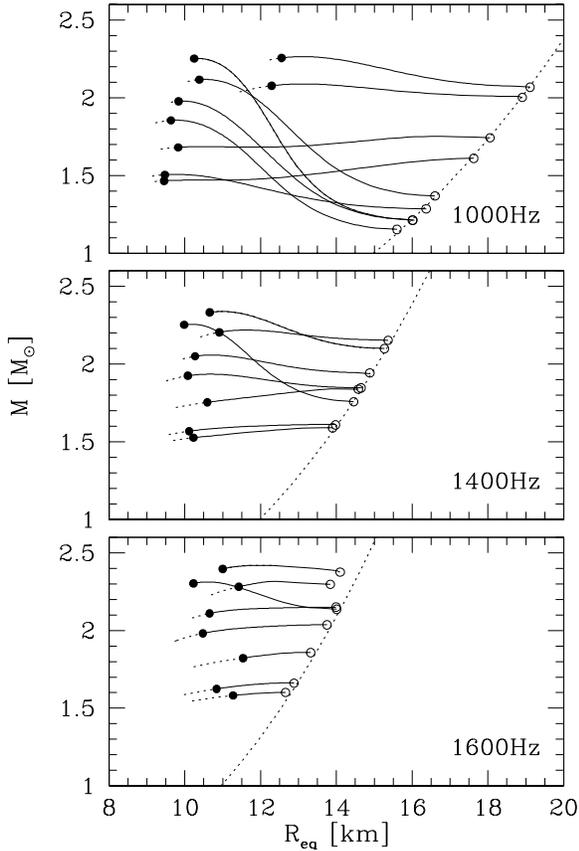}
\caption{$M$ vs  $R_{\rm eq}$ for stably rotating sub-millisecond
compact stars. Notations as in Fig.\ \ref{fig:MR-NS}. To indentify a
curve corresponding to a specific EOS from Table 1,
one has to use the sequence of open
circles (ordered in the same way as in this figure)
at the mass-shedding limit in Fig.\ \ref{fig:MR-NS}. }
\label{fig:MRsubms}
\end{figure}
Mass vs equatorial radius relations for very fast rotating
compact  stars  are presented in Fig.\ \ref{fig:MRsubms}. The
shape of the $M(R_{\rm eq})$ curves is more and more flat as
rotational frequency increases. For $f_{\rm rot}=1600~{\rm
Hz}$ the curves $M(R_{\rm eq})$ are almost horizontal. At that
frequency {\it  the mass for each EOS is quite well defined}.
Moreover,  curves for different EOSs are usually  well
separated. Both features are of a great practical importance.
In principle,  they are very useful for selecting "true EOS",
provided one detects a very fast  pulsar and is simultaneously
able to measure its mass.

A dotted curve in every panel of Fig.\ \ref{fig:MRsubms}
corresponds to the formula
\begin{equation}
M={4\pi^2 f^2\over G}R_{\rm eq}^3~,
\label{eq:M(R)f.ms}
\end{equation}
used for a panel  frequency (1000~Hz, 1400~Hz, 1600~Hz).
Notice that Eq.\ (3) is a special case of Eq.\
(\ref{eq:M(R)f.ms}).  Equation (\ref{eq:M(R)f.ms}) works very
well  in very  broad range of rotational frequencies and EOSs
(recently this formula has been tested by \citet{KLW2007} for
the frequency 716~Hz of  PSR J1748-2446ad.)
\section{Spin up by accretion}
\label{sect:accretion}
It is commonly believed that fast (millisecond) pulsars are
recycled old neutron stars,  spun up to kHz frequencies via a
long-time disk accretion in LMXBs. Such neutron stars have a
weak magnetic field ($B<10^{10}~$G), which does not affect the
accretion flow. Therefore, the accretion disk extends down to
the innermost stable circular orbit (ISCO).

In the present section we study some aspects of spin-up by
accretion from the ISCO. We  use the prescription given by
\citet{ZHG2002,ZHB2005}. Specific angular momentum
per unit baryon mass of a matter element  orbiting the neutron
star at the ISCO, $l_{\rm _{ISCO}}$, is calculated by solving
exact equations of the orbital motion of a particle in the
space-time produced by a rotating neutron star ( Appendix A of
 \citealt{ZHG2002}).

Consider accretion of an infinitesimal amount of baryon mass
${\rm d}M_{\rm B}$ onto a rotating neutron star. As the star
is assumed to  spin up in a quasi-stationary manner, accretion
of ${\rm d}M_{\rm B}$ leads to a new rigidly rotating
configuration of mass $M_{\rm B}+{\rm d}M_{\rm B}$ and angular
momentum $J+{\rm d}J$, with
\begin{equation}
{\rm d}J= x_{l}l_{\rm _{ISCO}} \;{\rm d}M_{\rm B}~.
\label{eq:lms}
\end{equation}
Here,   $x_l$ denotes the fraction of the angular momentum of
the matter element,  transferred to the star. The remaining
fraction, $1-x_l$, is assumed to be lost via radiation or
other dissipative  processes.
\begin{figure}

\includegraphics[width=\columnwidth]{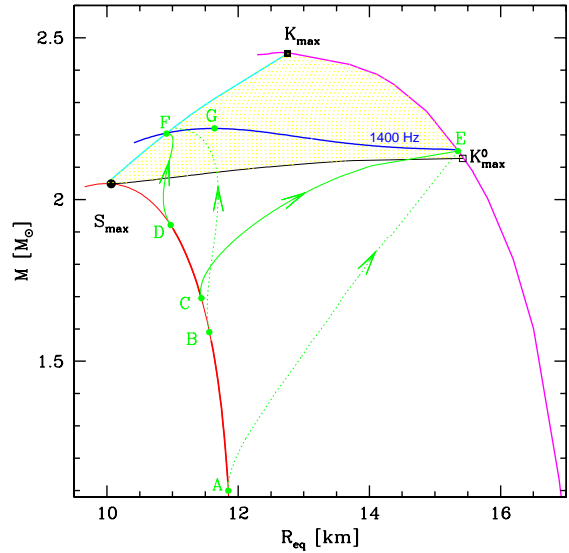}
\caption{Mass
vs radius relation for an accreting neutron star with DH EOS.
The star spins-up from $f=0$ to $f_{\rm rot}=1400$~Hz.
For further explanations see the text.}
\label{fig:MR1400}
\end{figure}
Let us consider two values of $x_l$: $x_l=1$ and
$x_l=0.5$. In Fig. \ref{fig:MR1400} we plot the curves
 $M(R_{\rm eq})$, corresponding to the spin-up from $f=0$
 to  (final) $f_{\rm rot}=1400$~Hz. Calculations are performed
  for the DH EOS. Point F on this curve corresponds to
  the onset of instability with respect to axi-symmetric
  oscillations (condition given by Eq.\ (\ref{eq:ax-sym.stab.line})).
  Point E is the Keplerian configuration at frequency 1400 Hz,
  while point  G  corresponds to
maximum mass along the curve with a fixed
rotation frequency $1400~$Hz.

Let us consider  several spin-up tracks, all terminating on
the 1400 Hz mass-radius line.  The curves starting at points
A,B,C and D are the tracks of  accreting neutron stars defined
by the Eq.\ ( \ref{eq:lms}),  for $x_l=1$ (solid line) and
$x_l=0.5$ (dotted line) for cases C,D,  and A,B, respectively.
In order to  reach the frequency 1400~Hz, one has  to start
with a non-rotating neutron star located  in the segment CD
(if $x_l=1$),  or AB (if $x_l=0.5$).  As one can see,  there
are bounds  on the initial mass of a non-rotating star, which
can be spun-up to a given frequency $f_{\rm rot}$ via disk
accretion. For $f_{\rm rot}=1400$~Hz $x_l=1$ the allowed mass
range for initial non-rotating star is
$1.7\msol<M_i<1.92\msol$.
\begin{figure}
\includegraphics[width=\columnwidth]{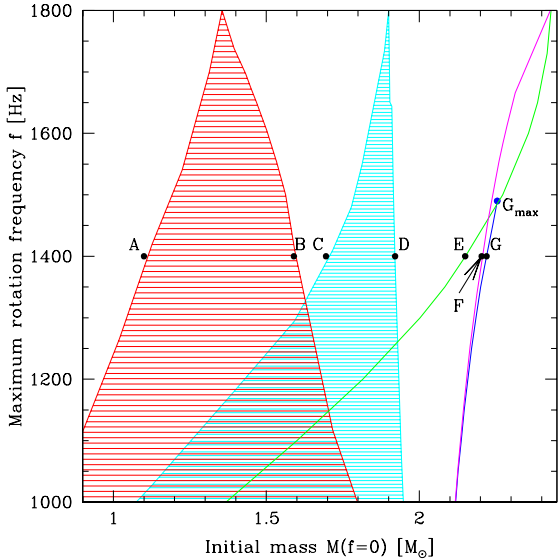}
\caption{(Color online) The bounds on the initial mass a
static neutron star (horizontal axis) to be spun-up up to a
given final frequency (vertical axis). Calculations for the DH
EOS. Labeled points (A, B, \ldots) correspond to the same
points in Fig. \ref{fig:MR1400}. The triangular region on  the
left (red) corresponds to $x_l=0.5$. Triangular region near
the center (cyan) is obtained for  $x_l=1$. The non-shaded
region on the right,  composed of a tilted triangle and a
lentil-like upper determines the range of  allowable masses of
final configurations rotating at a given frequency.}
\label{fig:fmacc}
\end{figure}

In Fig  \ref{fig:fmacc} we  plotted, for the DH EOS, the
bounds on the initial mass of a non-rotating neutron star, for
which the spin-up via accretion could reach a required
rotation frequency.  Results are presented for two values of
$x_l$. Shaded areas correspond to the  allowed a range of
initial masses of non-rotating star. Left "triangle" was
obtained for $x_l=0.5$, while the central shaded "triangle"
(with points C and D)  - for $x_l=1$.

The limits on the actual mass of rotating NS which was formed
by disk accretion onto an initial static star are given by the
three curves on the right in Fig.\ \ref{fig:fmacc}. These
curves pass by  three points E,F, and G in Fig.
\ref{fig:MR1400}. The curve passing by the point E (green) is
the bound resulting from the  Keplerian limit of a rotating
star. The (magenta) line passing by the point  F is the
boundary resulting from the instability with respect to
axi-symmetric perturbations. Finally, curve passing through
the point G is a locus of maxima on the mass-radius curve at a
fixed final rotation frequency: point G is obtained for 1400
Hz. As the frequency of the final configuration increases, the
mass at the Keplerian limit increases more rapidly, than that
defined by the onset of the axi-symmetric instability at the
maximum mass (at 1400 Hz: points F and G,  respectively). For
very high frequencies the maximum mass of the stars rotating
at a fixed frequency is given by the value for Keplerian
limit. In the considered case of the  DH EOS, the point G
disappears  at frequency $\simeq 1500$~Hz (point ${\rm G}_{\rm
max}$). For faster rotation the curve $M(R_{\rm eq})$ is
monotonous (no maximum between the both ends). However, the
range of masses of stars rotating at so high frequency is very
narrow. For $f_{\rm rot}
> 1400$~Hz it is smaller than $0.1\msol$ (see also discussion
of Fig.\ref{fig:MRsubms} in Sect.\ \ref{sect:sub-ms}).

\section{Discussion and conclusions}
\label{sect:discuss}
Let us summarize main results reported in  the present paper,
and obtained using a large set of theoretical EOSs.  The
$M(R_{\rm eq})$ curve for $f\gtrsim 1400$~Hz (rotation period
$\lesssim 0.7~$ms)  is flat. Therefore, at such frequencies,
for any  given EOS the mass of a stably rotating compact star
is quite well defined. Conversely, a measured mass of a
compact star rotating at $f\gtrsim 1400$ Hz will allow us to
unveil the actual EOS of dense matter. The "Newtonian" formula
for the Keplerian frequency reproduces  surprisingly well
precise 2-D simulations and sets a firm upper limit on $R_{\rm
eq}$ for a given $f$. Finally, observation of $f\gtrsim 1200$
Hz sets stringent limits on the initial mass of a non-rotating
star which was spun up to this frequency via accretion from a
disk.

In the present paper we limited ourselves to {\it hadronic
stars}, built exclusively or predominantly of hadrons. We did
not discuss our results obtained for hypothetical compact
stars built of a self-bound quark matter, called {\it quark
stars} or {\it strange stars}. Some of results for quark stars
with and without crust were briefly reported in Ref.\
\citet{BHZ2007}. Generally, most of general features and
relations obtained for hadronic stars hold also for quark
stars. However, is some cases one notices systematic
difference between rapidly rotating hadronic and quark stars.
Our results will be described in a forthcoming publication.
%
\section*{Acknowledgments}
 This work was partially
supported by the Polish MNiSW grant no. N203.006.32/0450
and by the LEA
Astrophysics Poland-France
(Astro-PF) program. MB was
also partially supported by the Marie Curie Intra-european
Fellowship MEIF-CT-2005-023644.

\bibliographystyle{elsart-harv}

\end{document}